\documentclass[twocolumn,A4]{article}
\usepackage[dvips]{graphics}
\usepackage{setspace}
\usepackage{color}
\definecolor{gold}{rgb}{0.85,0.66,0}
\definecolor{dblue}{rgb}{0,0,0.5}
\begin{document}
\onecolumn
\begin{center}
{\bf{\Large {\textcolor{gold}{Effect of localizing groups on electron 
transport through single conjugated molecules}}}}\\
~\\
{\textcolor{dblue}{Santanu K. Maiti}}$^{1,2,*}$ \\
~\\
{\em $^1$Theoretical Condensed Matter Physics Division,
Saha Institute of Nuclear Physics, \\
1/AF, Bidhannagar, Kolkata-700 064, India \\
$^2$Department of Physics, Narasinha Dutt College,
129, Belilious Road, Howrah-711 101, India} \\
~\\
{\bf Abstract}
\end{center}
Electron transport properties through single conjugated molecules 
sandwiched between two non-superconducting electrodes are studied 
by the use of Green's function technique. Based on the tight-binding 
model, we do parametric calculations to characterize the electron 
transport through such molecular bridges. The electron transport 
properties are significantly influenced by (a) the existence of 
localizing groups in these conjugated molecules and (b) the molecule 
to electrode coupling strength, and, here we focus our results in 
these two aspects.

\vskip 1cm
\begin{flushleft}
{\bf PACS No.}: 73.23.-b; 81.07.Nb; 85.65.+h \\
~\\
{\bf Keywords}: Conjugated molecules; Localizing groups; Conductance;
$I$-$V$ characteristic.
\end{flushleft}
\vskip 4.75in
\noindent
{\bf ~$^*$Corresponding Author}: Santanu K. Maiti

Electronic mail: santanu.maiti@saha.ac.in
\newpage
\twocolumn

\section{Introduction}
Molecular electronics and transport have attracted much more attention
since molecules constitute promising building blocks for future generation
of nanoelectronic devices. Following experimental developments, theory 
can play a major role in understanding the new mechanisms of conductance. 
The single-molecule electronics plays a significant role in designing 
and developing the future nanoelectronic circuits, but, the goal of 
developing a reliable molecular-electronics technology is still over 
the horizon and many key problems, such as device stability, 
reproducibility and the control of single-molecule transport need to 
be solved. Electronic transport through molecules was first studied 
theoretically in $1974$~\cite{aviram}. Then lot of 
experiments~\cite{tali,metz,fish,reed1,reed2} have been performed 
through molecules placed between two metallic electrodes with few 
nanometer separation. The operation of such two-terminal devices is 
due to an applied bias. Current passing across the junction is strongly 
nonlinear function of applied bias voltage and its detailed description 
is a very complex problem. The complete knowledge of the conduction 
mechanism in this scale is not well understood even today. In many 
molecular devices, electronic transport is dominated by conduction
through broadened HOMO or LUMO states. In contrast here we find that the
transport through single conjugated molecules can be controlled very 
sensitively by introducing the localizing groups in these molecules. This 
sensitivity opens up new possibilities for novel single-molecule sensors.
Electron conduction through molecules strongly depends on (a) the 
delocalization of the molecular electronic orbitals and (b) their coupling 
strength to the two electrodes. In a very recent experiment, Tali Dadosh
{\em et al.}~\cite{tali} have measured conductance of single conjugated
molecules and predicted that the existence of localizing groups in a
conjugated molecule suppresses the electrical conduction through the
molecule. These results motivate us to study the electron transport 
through such conjugated molecules.

The aim of the present paper is to reproduce an analytic approach 
based on the tight-binding model to investigate the electron transport 
properties for the model of single conjugated molecules taken in their
experiment~\cite{tali}. Several {\em ab initio} methods are used for the
calculation of conductance~\cite{ven,yal,xue,tay,der,dam}, yet it is
needed the simple parametric approaches~\cite{muj1,muj2,sam,hjo,baer1,
baer2,baer3,walc1,walc2} for this calculation. The parametric study is 
motivated by the fact that it is much more flexible than that of the 
{\em ab initio} theories since the later theories are computationally 
very expensive and here we focus our attention on the qualitative 
effects rather than the quantitative ones. This is why we restrict our 
calculations on the simple analytical formulation of the transport problem.

The scheme of the paper is as follow. In Section $2$, we give a very 
brief description for the calculation of transmission probability and 
current through a finite size conductor sandwiched between two 
one-dimensional ($1$D) metallic electrodes. Section $3$ focuses the 
results of conductance-energy ($g$-$E$) and current-voltage ($I$-$V$) 
characteristics for the single conjugated molecules and study the 
effects of localizing groups in the above mentioned quantities. 
Finally, we summarize our results in Section $4$.

\section{A glimpse onto the theoretical formulation}

Here we describe very briefly about the methodology for the calculation 
of transmission probability ($T$), conductance ($g$) and current ($I$) 
through a finite size conducting system attached to two semi-infinite 
metallic electrodes by using the Green's function technique.

\begin{figure}[ht]
{\centering \resizebox*{7.5cm}{1.5cm}{\includegraphics{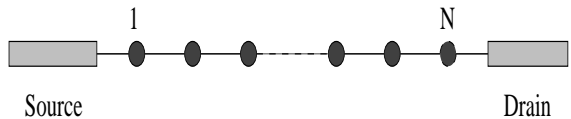}}\par}
\caption{Schematic view of a $1$D conductor with $N$ number of atomic 
sites (filled circles) attached to two electrodes through the sites $1$ 
and $N$, respectively.}
\label{dot}
\end{figure}

Let us first consider a $1$D conductor with $N$ number of atomic sites 
(array of filled circles) connected to two semi-infinite electrodes, 
namely, source and drain, as presented in Fig.~\ref{dot}. The conducting 
system in between
the two electrodes can be an array of few quantum dots, or a single 
molecule, or an array of few molecules, etc. At low voltages and 
temperatures, the conductance of the conductor can be written by 
using the Landauer conductance formula,
\begin{equation}
g=\frac{2e^2}{h}T
\label{land}
\end{equation}
where $g$ is the conductance and $T$ is the transmission probability 
of an electron through the conductor. The transmission probability can 
be expressed in terms of the Green's function of the conductor and the 
coupling of the conductor to the two electrodes by the expression,
\begin{equation}
T={\mbox{Tr}} \left[\Gamma_S G_C^r \Gamma_D G_C^a\right]
\label{trans1}
\end{equation}
where $G_C^r$ and $G_C^a$ are respectively the retarded and advanced Green's
function of the conductor. $\Gamma_S$ and $\Gamma_D$ are the coupling terms
of the conductor due to the coupling to the source and drain, respectively.
For the complete system, i.e., the conductor and the two electrodes, the
Green's function is defined as,
\begin{equation}
G=\left(\epsilon-H\right)^{-1}
\end{equation}
where $\epsilon=E+i\eta$. $E$ is the injecting energy of the source electron
and $\eta$ is a very small number which can be put as zero in the limiting
approximation. The above Green's function corresponds to the inversion of an
infinite matrix which consists of the finite conductor and two semi-infinite
electrodes. It can be partitioned into different sub-matrices those 
correspond to the individual sub-systems.

The effective Green's function for the conductor can be written as,
\begin{equation}
G_C=\left(\epsilon-H_C-\Sigma_S-\Sigma_D\right)^{-1}
\end{equation}
where $H_C$ is the Hamiltonian for the conductor sandwiched between the two
electrodes. The single band tight-binding Hamiltonian for the conductor 
within the non-interacting picture can be written in the following form,
\begin{equation}
H_C=\sum_i \epsilon_i c_i^{\dagger} c_i + \sum_{<ij>}t 
\left(c_i^{\dagger}c_j + c_j^{\dagger}c_i \right)
\label{hamil1}
\end{equation}
where $c_i^{\dagger}$ ($c_i$) is the creation (annihilation) operator of an
electron at site $i$, $\epsilon_i$'s are the site energies and $t$ is the
nearest-neighbor hopping integral. Here $\Sigma_S=h_{SC}^{\dagger} g_S h_{SC}$
and $\Sigma_D=h_{DC} g_D h_{DC}^{\dagger}$ are the self-energy terms due to
the two electrodes. $g_S$ and $g_D$ are respectively the Green's function for
the source and drain. $h_{SC}$ and $h_{DC}$ are the coupling matrices and they
will be non-zero only for the adjacent points in the conductor, $1$ and $N$ as
shown in Fig.~\ref{dot}, and the electrodes respectively. The coupling terms
$\Gamma_S$ and $\Gamma_D$ for the conductor can be calculated through the
expression,
\begin{equation}
\Gamma_{\{S,D\}}=i\left[\Sigma_{\{S,D\}}^r-\Sigma_{\{S,D\}}^a\right]
\end{equation}
where $\Sigma_{\{S,D\}}^r$ and $\Sigma_{\{S,D\}}^a$ are the retarded and
advanced self-energies, respectively, and they are conjugate to each
other. Datta {\em et al.}~\cite{tian} have shown that the self-energies
can be expressed like,
\begin{equation}
\Sigma_{\{S,D\}}^r=\Lambda_{\{S,D\}}-i \Delta_{\{S,D\}}
\end{equation}
where $\Lambda_{\{S,D\}}$ are the real parts of the self-energies which
correspond to the shift of the energy eigenstates of the conductor and the
imaginary parts $\Delta_{\{S,D\}}$ of the self-energies represent the
broadening of these energy levels. This broadening is much larger than the 
thermal broadening and this is why we restrict our all calculations only 
at absolute zero temperature. By doing some simple algebra these real and 
imaginary parts of self-energies can also be determined in terms of
coupling strength ($\tau_{\{S,D\}}$) between the conductor and two 
electrodes, injection energy ($E$) of the transmitting electron, site 
energy ($\epsilon_0$) of the electrodes and hopping strength ($v$) between 
nearest-neighbor sites in the electrodes. Thus the coupling terms 
$\Gamma_S$ and $\Gamma_D$ can be written in terms of the retarded 
self-energy as,
\begin{equation}
\Gamma_{\{S,D\}}=-2 {\mbox{Im}} \left[\Sigma_{\{S,D\}}^r\right]
\end{equation}
Now all the information regarding the conductor to electrode coupling is
included into the two self energies as stated above and is analyzed 
through the use of Newns-Anderson chemisorption theory~\cite{muj1,muj2}. 
The detailed description of this theory is obtained in these two 
references.

By calculating the self-energies, the coupling terms $\Gamma_S$ and
$\Gamma_D$ can be easily obtained and then the transmission probability 
$T$ can be computed from the expression as mentioned in Eq.~\ref{trans1}.

Since the coupling matrices $h_{SC}$ and $h_{DC}$ are non-zero only for
the adjacent points in the conductor, $1$ and $N$ as shown in Fig.~\ref{dot},
the transmission probability becomes,
\begin{equation}
T(E,V)=4\Delta_{11}^S(E,V) \Delta_{NN}^D(E,V)|G_{1N}(E,V)|^2
\label{trans2}
\end{equation}
The current passing through the conductor is depicted as a single-electron
scattering process between the two reservoirs of charge carriers. The 
current-voltage relation is evaluated from the following
expression~\cite{datta},
\begin{equation}
I(V)=\frac{e}{\pi \hbar}\int \limits_{E_F-eV/2}^{E_F+eV/2} T(E,V) dE
\end{equation}
where $E_F$ is the equilibrium Fermi energy. For the sake of simplicity, 
here we assume that the entire voltage is dropped across the 
conductor-electrode interfaces and this assumption does not significantly 
change the qualitative behaviors of the $I$-$V$ characteristics. Using the 
expression of $T(E,V)$ as in Eq.~\ref{trans2} the final form of $I(V)$ 
becomes,
\begin{eqnarray}
I(V) &=& \frac{4e}{\pi \hbar}\int \limits_{E_F-eV/2}^{E_F+eV/2}
\Delta_{11}^S(E,V) \Delta_{NN}^D(E,V) \nonumber \\
& & \times~ |G_{1N}(E,V)|^2 dE
\label{curr}
\end{eqnarray}
Eq.~\ref{land}, Eq.~\ref{trans2} and Eq.~\ref{curr} are the final working
formulae for the calculation of conductance $g$ and current-voltage
characteristics, respectively, for any finite size conductor sandwiched
between two electrodes.

With the help of the above formulation, we shall describe the electron 
transport properties through some conjugated molecules (Fig.~\ref{conju}). 
For the sake of simplicity throughout this article we use the unit $c=e=h=1$.

\section{Results and discussion}

This section focuses the conductance-energy ($g$-$E$) and current-voltage 
($I$-$V$) characteristics of three short single conjugated molecules. 
These molecules are specified as: 
\begin{figure}[ht]
{\centering \resizebox*{8cm}{6cm}{\includegraphics{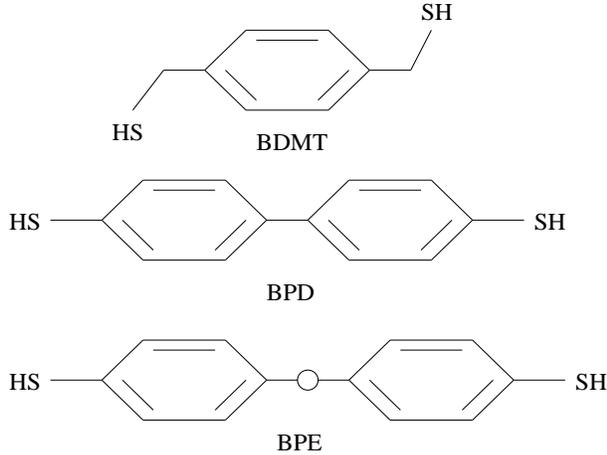}}\par}
\caption{Structures of the three molecules: $1$,$4$-benzenedimethanethiol
(BDMT), $4$,$4^{\prime}$-biphenyldithiol (BPD) and
bis-($4$-mercaptophenyl)-ether (BPE) those are attached to two electrodes 
by thiol (S-H) groups.}
\label{conju}
\end{figure}
$1$,$4$-benzenedimethanethiol (BDMT), in which the molecular conjugation 
is broken near the contacts by a methylene group; 
$4$,$4^{\prime}$-biphenyldithiol (BPD), a fully conjugated molecule; and 
bis-($4$-mercaptophenyl)-ether (BPE), where the molecular conjugation is 
broken by an oxygen atom at the center. The schematic representations of 
these three molecules, with thiol groups at the two extreme ends of each 
molecules, are shown in Fig.~\ref{conju}. These molecules are contacted 
to the two semi-infinite $1$D electrodes by thiol (S-H) groups via single 
channels (same as shown schematically in Fig.~\ref{dot}). In actual 
experimental arrangement, two electrodes are constructed by using gold 
(Au) substance and molecule attached to the electrodes by thiol (S-H) 
\begin{figure}[ht]
{\centering \resizebox*{7cm}{10cm}{\includegraphics{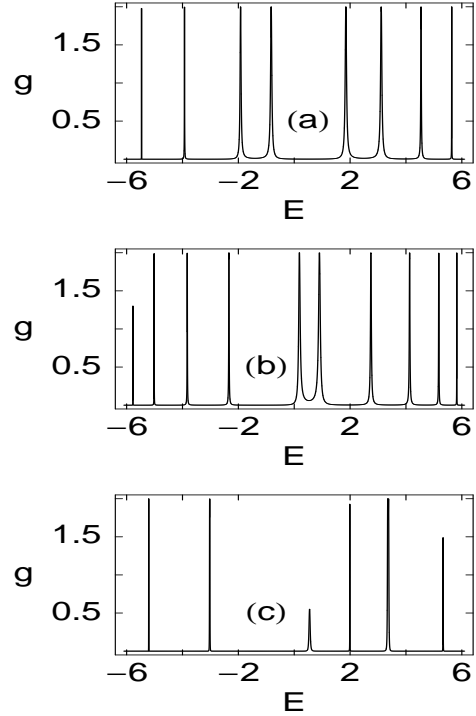}}\par}
\caption{Conductance $g$ as a function of the injecting electron energy 
$E$ in the weak-coupling limit, where (a), (b) and (c) are respectively 
for the BDMT, BPD and BPE molecules.}
\label{condlow}
\end{figure}
groups in the chemisorption technique where hydrogen (H) atoms remove 
and sulfur (S) atoms reside. The electron transport through such conjugated 
molecules significantly influenced by the presence of localizing groups in 
the molecules and the molecule-to-electrode coupling strength. Here, we 
shall investigate our results in these aspects. Throughout the article, 
we discuss the results in two limiting regimes depending on the coupling 
strength of the molecule to the electrodes. One is defined as 
$\tau_{\{S,D\}}<<t$, the so-called weak-coupling limit. The other one is 
$\tau_{\{S,D\}}\sim t$, the so-called strong-coupling limit. The 
parameters $\tau_S$ and $\tau_D$ correspond to the coupling of the 
molecules to the source and drain, respectively. The values of the 
different parameters used in our calculations in these two limiting 
regimes are assigned as: $\tau_S=\tau_D=0.5$; $t=2.5$ (weak coupling) 
and $\tau_S=\tau_D=2$; $t=2.5$ (strong-coupling). For the side attached 
electrodes the on-site energy ($\epsilon_0$) and the nearest-neighbor 
hopping strength ($v$) are fixed to $0$ and $4$, respectively. The Fermi 
energy $E_F$ is set at $0$.

In Fig.~\ref{condlow} we display conductance ($g$) as function of injecting 
electron energy ($E$) for the three molecular bridge systems in the limit of
weak molecular coupling. Figures~\ref{condlow}(a), (b) and (c) correspond to
\begin{figure}[ht]
{\centering \resizebox*{7cm}{10cm}{\includegraphics{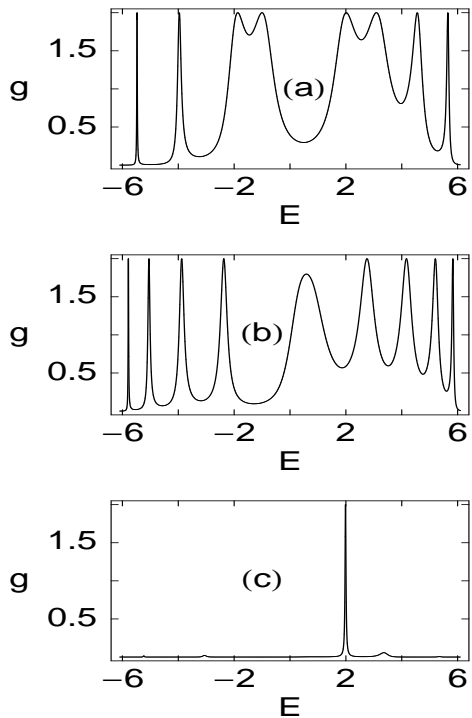}}\par}
\caption{Conductance $g$ as a function of the injecting electron energy 
$E$ in the strong molecule-to-electrode coupling limit, where (a), (b) 
and (c) are respectively for the BDMT, BPD and BPE molecules.}
\label{condhigh}
\end{figure}
the results for the bridges with BDMT, BPD and BPE molecules, respectively. 
Conductance shows very sharp resonant peaks for some particular energy 
values, while in almost all other cases it drops to zero. These resonant
peaks are associated with the energy eigenstates of the individual molecules 
that bridges the two reservoirs. Therefore, the conductance spectrum 
manifests itself the electronic structure of the molecule. At resonances, 
the conductance ($g$) achieves the value $2$, and accordingly, the 
transmission probability ($T$) goes to unity since we have the relation
$g=2T$ from the Landauer conductance formula with $e=h=1$ in our present 
formulation. For the bridges with BDMT and BPD molecules, we see that the 
resonant peaks have very narrow widths, while for the bridge with BPE 
molecule the width of the peaks is almost zero. Thus, fine tuning in the 
energy scale is necessary to get the electron conduction through these 
bridges, specially in the case of BPE molecule for the weak-coupling 
limit. The most significant result is that, the BPD molecule conducts 
\begin{figure}[ht]
{\centering \resizebox*{7.7cm}{5cm}{\includegraphics{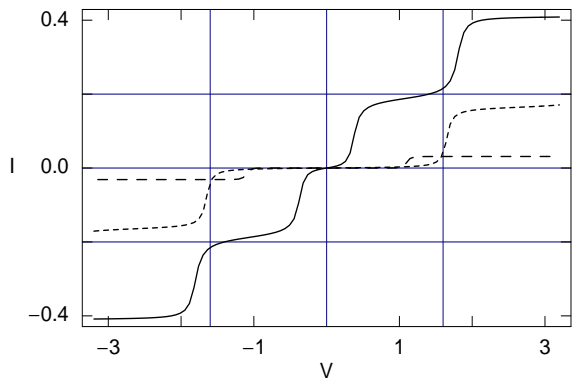}}\par}
\caption{Current $I$ as a function of the applied bias voltage $V$ in 
the weak molecule-to-electrode coupling, where the solid, dotted 
and dashed lines are respectively for the BPD, BDMT and BPE molecules.}
\label{currlow}
\end{figure}
electron across the zero energy value, while the other two conduct beyond 
some critical energy values (see Fig.~\ref{condlow}). Therefore, we can 
tune the electron conduction through the molecular bridge in a very 
controllable way.

In the strong molecular coupling limit, the resonant peaks in the
conductance spectra get substantial widths as shown in Fig.~\ref{condhigh}. 
This enhancement of the resonant widths is due to the broadening of the 
molecular energy eigenstates caused by the coupling of the molecules to 
the side attached electrodes in the strong-coupling limit, where the 
contribution comes from the imaginary parts of the 
self-energies~\cite{datta}. Though for the molecular bridges with 
BDMT and BPD molecules the resonant peaks get substantial widths, but, 
for the bridge with BPE molecule, the increment of the widths is very 
small. For this BPE molecule since the increment of the width of the 
resonant peak across the energy $E=2$ is comparatively higher than 
for the other energy values, we observe only one peak across this 
energy value ($E=2$, Fig.~\ref{condhigh}(c)). Thus, for this molecular 
bridge electron conduction takes place across a particular energy value, 
while in all other energies no electron conduction takes place. This 
aspect may be used to describe switching action in an electronic circuit.

Thus we see that the electron conduction strongly depends on the molecule 
itself and also on the strength of the molecular coupling to the side 
attached electrodes. The behavior of electron transfer through the 
molecular junction becomes much more clearly observed from the 
current-voltage characteristics. Current passing through the molecular 
system is computed from the integration procedure of the transmission 
function $T$. The nature of the transmission function is exactly similar 
to that of the conductance spectrum since $g=2T$ (from the Landauer 
formula), differ only in magnitude by the factor $2$.
\begin{figure}[ht]
{\centering \resizebox*{7.7cm}{5cm}{\includegraphics{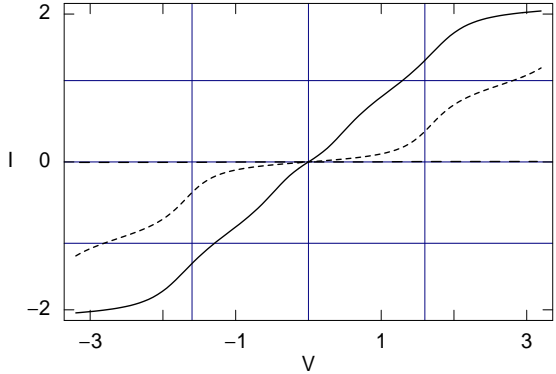}}\par}
\caption{Current $I$ as a function of the applied bias voltage $V$ in 
the strong molecule-to-electrode coupling, where the solid, dotted and 
dashed curves are respectively for the BPD, BDMT and BPE molecules.}
\label{currhigh}
\end{figure}
In Fig.~\ref{currlow}, we plot the current-voltage characteristics of 
these three molecular bridges in the weak-coupling limit. The solid, 
dotted and dashed curves correspond to the results for the molecular 
bridges with BPD, BDMT and BPE molecules, respectively. The current 
shows staircase-like structure with fine steps as a function of the 
applied bias voltage. This is due to the discreteness of molecular 
resonances as shown in Fig.~\ref{condlow}. With the increase of the 
bias voltage, the electrochemical potentials on the electrodes are 
shifted and eventually cross one of the molecular energy level. 
Accordingly, a current channel is opened and a jump in the $I$-$V$ 
curve appears. The significant observation is that, for the molecular 
bridge with BPD molecule (free from localizing group), the current 
amplitude is much higher (see solid curve of Fig.~\ref{currlow}) 
compared two the other two bridges. This is due to the fact that the 
localizing groups (both in BDMT and BPE molecules) interfere with the 
conjugated aromatic systems and suppress the overall conductance 
through the molecules. On the other hand, the another important 
feature is that, in purely conjugate molecule (BPD) the electron 
conduction takes place as long as the bias voltage is applied, while
for the other two molecules it appears beyond some finite values of 
$V$. This behavior gives a key idea in the fabrication of molecular 
devices.

The shape and height of these current steps depend on the width of the 
molecular resonances. With the increase of molecule-to-electrode coupling 
strength, current gets a continuous variation with the applied bias voltage 
and achieves much higher values (compared to the current amplitude in
the weak coupling case), as plotted in Fig.~\ref{currhigh}, where 
the solid, dotted and dashed curves correspond to the same meaning as in 
Fig.~\ref{currlow}. In this strong molecular coupling limit, the current 
amplitude for the molecular bridge with BPE molecule is negligibly small 
compared to the other two bridges and the other features are also similar 
to the case of weak molecular coupling limit. 

\section{Concluding remarks}

In summary, we have studied electron transport, at absolute zero 
temperature, through three short single conjugated molecules based 
on the tight-binding framework. We have used parametric approach, 
since we are interested only on the qualitative behaviors instead 
of the quantitative ones, rather than the {\em ab initio} theories 
since the later theories are computationally too expensive. This 
technique can be used to study the electronic transport in any 
complicated molecular system. 

Electronic transport is significantly affected by (a) the molecule 
itself and (b) molecule-to-electrode coupling strength and in this 
article we have studied our results in these aspects. In the weak-coupling 
limit conductance shows sharp resonant peaks, while these peaks get 
broadened in the limit of strong molecular coupling. These results 
predict that by tuning the molecular coupling strength one can control 
the electron conduction very sensitively through the molecular bridges. 
In the study of current we have seen that the current shows step-like 
behavior with sharp steps in the weak molecular coupling, while it becomes 
continuous in the strong-coupling limit as a function of applied bias 
voltage. Both for the two limiting cases our results have clearly 
described that the localizing groups suppress the current amplitude 
in large amount compared to the current amplitude in case of purely 
conjugate molecule. Another significant observation is that the 
threshold bias voltage of electron conduction across a molecular bridge 
strongly depends on the molecule itself. These results provide key ideas 
for fabrication of different molecular devices, especially in the 
fabrication of molecular switches.

Some assumptions have been taken into account for this present study.
More studies are expected to take the Schottky effect which comes from the 
charge transfer across the molecule-electrode interfaces, the static Stark 
effect, which is taken into account for the modification of the electronic 
structure of the bridge system due to the applied bias voltage (essential 
especially for higher voltages). However, all these effects can be included 
into our framework by a simple generalization of the presented formalism. 
In this article we have also neglected the effects of inelastic scattering 
processes and electron-electron correlation to characterize the electron 
transport through such bridges.

\end{document}